\documentstyle[12pt]{article}
\begin{document}
{\large

\begin{center}
{\Large \bf
{GENERAL-RELATIVISTIC MODEL OF }\\
\vskip 0.3cm
{MAGNETICALLY DRIVEN JET}
}\\
\vskip 1.8cm
Andria~D.~Rogava\\
\vskip 0.3cm
{\it Department of Theoretical Astrophysics,  Abastumani  Astrophysical
Observatory, Republic of Georgia and Department of Physics, Tbilisi State
University, Republic of Georgia}\\
\vskip 0.8cm
George~ R.~Khujadze\\
\vskip 0.3cm
{\it Department of Theoretical Astrophysics,  Abastumani  Astrophysical
Observatory, Republic of Georgia}
\end{center}
\vskip 1.5cm
\begin{flushleft}
{Running Head: General-Relativistic Jet}\\
\vskip 2.4cm
{Mailing Address:}\\
{~~~~~~~~~~~~~~~~~~~~~~A.~D.~Rogava,}\\
{~~~~~~~~~~~~~~~~~~~~~~Department of Theoretical Astrophysics,}\\
{~~~~~~~~~~~~~~~~~~~~~~Abastumani  Astrophysical Observatory,}\\
{~~~~~~~~~~~~~~~~~~~~~~Kazbegi str. $N.~2^{a}$,}\\
{~~~~~~~~~~~~~~~~~~~~~~Tbilisi 380060,}\\
{~~~~~~~~~~~~~~~~~~~~~~Republic of Georgia.}\\
{~~~~~~~~~~~~~~~~~~~~~~Telephone: 7--8832--383510}\\
{~~~~~~~~~~~~~~~~~~~~~~E-mail: andro@dtapha.kheta.ge}\\
\end{flushleft}
\newpage

\section{Itroduction}

Powerful bipolar outflows appear to be common features of the wide class of
astrophysical objects including protostars, young main-sequence stars,
galactic X-ray sources, Active Galactic Nuclei (AGN), and quasars [1]. In
the majority of these cases there exists more or less convincing
observational evidence about the presence of large-scale magnetic fields in
these objects. Ordered magnetic fields should have the crucial role in
giving rise to such outflows or jets. Namely, they may be responsible for
the collimation of the jets and/or acceleration of matter up to relativistic
velocities in them [2].

High compactness and huge energy output, which characterizes some (maybe,
the most interesting) concrete kinds of objects featuring bipolar (or,
unipolar) outflows (such as AGNs and quasars), may be explained if one assumes
that they are produced in result of accretion onto the rotating supermassive
black hole. It seems evident that a consideration of the innermost part of
such outflows, in the close neighbourhood of the black hole must be performed
in the framework of general-relativistic magnetohydrodynamics (MHD). Such
treatment enables us to take into account properly the influence of strong
gravitational and electromagnetic fields onto the structure of the jet.
Since $3+1$ formulation of black hole electrodynamics [3--5] is the most
convenient mathematical apparatus for such purposes, in the present study
we shall establish our consideration on it.

It must be noted from the very beginning that the method, which is developed
in this paper is the generalization of the advanced methodology used in the
number of recent references for {\it non-relativistic} [6--9], and for {\it
special-relativistic} [10--11] magnetically driven jets and winds. By means
of the method we derive the set of equations for general-relativistic jets
from the equations of the ideal $3+1$ MHD. Under certain simplifying
assumptions we obtain the simple, representative solution of these
equations. The features of the solution are discussed and compared with the
one, which has been found in [8].

\section{Governing equations}

In the forthcoming analysis we shall use the following notations:~(a)
greek indices will range over $t,r,\theta,\phi$ and represent
space-time coordinates, components, etc.; (b)~Latin indices will
range over $r,\theta,\phi$ and represent coordinates in
three-dimensional "absolute" space. We will approve the spacelike signature
convention $(-~+~+~+)$ and units in which the gravitational constant and
the speed of light are equal to one.

The rotation of a central object (i.e. a rapidly rotating Kerr black hole)
introduces off-diagonal terms $g_{t\phi}$ in the metric so that the
space-time generated by the rotating body is represented by the metric:
$$
ds^{2}=g_{tt}dt^{2}+2g_{t\phi}dtd{\phi}+g_{\phi\phi}
d{\phi}^{2}+g_{rr}dr^2+g_{\theta\theta}d{\theta}^{2}, \eqno(1)
$$
with the metric coefficients independent of $t$ and $\phi$.

In 3+1  notations (1) may be rewritten as [3--5]:
$$
ds^{2}=-{\alpha}^{2}dt^{2}+{\gamma}_{ik}(dx^{i}+{\beta}^{i}dt)
(dx^{k}+{\beta}^{k}dt), \eqno(2)
$$
where $\alpha$ is the so-called {\it lapse function} defined as:
$$
{\alpha}^{2}{\equiv}{{g_{t\phi}^{2}-g_{tt}g_{\phi\phi}}
\over g_{\phi\phi}}, \eqno(3)
$$
${\gamma}_{ik}$ is the three-dimensional ``absolute" space metric tensor
(with diagonal nonzero components $g_{ii}{\equiv}{\gamma}_{ii}$) and
{$\vec \beta$} is the spatial (three-dimensional) vector with components
$$
{\beta}^{i}{\equiv}{\left(0,~0,~{{g_{t\phi}}\over
g_{\phi\phi}}\right)};~~~{\beta}_{i}={\gamma}_{ik}{\beta}^{k}. \eqno(4)
$$

Note that the Kerr metric is the subclass of the general metric (1).
In particular,
$$
g_{tt}=-{\left(1-{{2Mr}\over{{\Sigma}}}\right)},   \eqno(5)
$$
$$
g_{t\phi}=-{{2aMrsin^2{\theta}}\over{{\Sigma}}},  \eqno(6)
$$
$$
g_{{\phi}{\phi}}={{Asin^2{\theta}}\over{{\Sigma}}},  \eqno(7)
$$
$$
g_{rr}={{\Sigma}\over{\Delta}},   \eqno(8)
$$
$$
g_{{\theta}{\theta}}={\Sigma};     \eqno(9)
$$
where $a{\equiv}J/M$ is the specific angular momentum of the black hole
per its unit mass, and
$$
\Sigma{\equiv}r^2+a^2cos^2{\theta},  \eqno(10)
$$
$$
\Delta{\equiv}r^2-2Mr+a^2, \eqno(11)
$$
$$
A{\equiv}(r^2+a^2)^2-a^2{\Delta}sin^2{\theta}.   \eqno(12)
$$

In the present study we consider jets as being axisymmetric
(${\partial}_{\phi}=0$) and stationary (${\partial}_t=0$). We, also,
neglect dissipative effects. The basic equations are those of
the general-relativistic MHD written in $3+1$ formalism:
$$
{\nabla}[{\Gamma}mn(\alpha{\vec v}-{\vec {\beta}})]=0,    \eqno(13)
$$
$$
{\nabla}{\times}(\alpha\vec B)=4{\pi}{\alpha}{\vec J} -
{\cal L}_{\vec {\beta}}{\vec E},     \eqno(14)
$$
$$
{\vec E}+{\vec v}{\times}{\vec B}=0,  \eqno(15)
$$
$$
{\nabla}{\times}({\alpha}{\vec E})={\cal L}_{\vec {\beta}}{\vec B}, \eqno(16)
$$
$$
{\nabla}{\vec B}=0,    \eqno(17)
$$
$$
{{d{\varepsilon}}\over{d{\tau}}}=-{1\over{{\alpha}^2}}{\nabla}({\alpha}^2
{\vec S})-{\sigma}_{ik}T^{ik},  \eqno(18)
$$
$$
{{d{\vec S}}\over{d{\tau}}}={\varepsilon}{\vec g}+{\bf H}{\cdot}{\vec S}-
{{1}\over{{\alpha}}}{\nabla}({\alpha}{\bf T}).  \eqno(19)
$$

In these equations we use the following notations:
$$
{\Gamma}=[1-({\vec v}{\cdot}{\vec v})]^{-1/2},    \eqno(20)
$$
$$
{{d}\over{d{\tau}}}{\equiv}{{1}\over{{\alpha}}}[{\partial}_{t}-
({\vec {\beta}}{\cdot}{\nabla})],     \eqno(21)
$$
$$
{\vec g}{\equiv}-{{1}\over{{\alpha}}}{\nabla}({\alpha}),    \eqno(22)
$$
$$
H_{ik}{\equiv}{{1}\over{{\alpha}}}{\beta}_{k:i},    \eqno(23)
$$
$$
{\sigma}_{ik}{\equiv}-{1 \over 2}(H_{ki}+H_{ik}); \eqno(24)
$$
and all vector and tensor quantities are defined in the three-dimensional
``absolute" space with the metric ${\gamma}_{ik}$.

Note also that in these equations $n$ is the proper baryon number
density, while
${\vec E}$, ${\vec B}$, and ${\vec J}$ are the vectors of the electric field,
the magnetic field and the current density, respectively. The
``Lie derivative" of a vector ${\vec A}$ along the vector ${\vec {\beta}}$
is defined in the following way:
$$
{\cal L}_{{\vec {\beta}}}{\vec A}{\equiv}({\vec A},{\nabla}){\vec {\beta}}-
({\vec {\beta}},{\nabla}){\vec A},   \eqno(25)
$$

and ${\vec v}$ is 3-velocity, related to the spatial components of
matter 4-velocity $u^{\alpha}{\equiv}(u^{t}, u^{i})$ via the expressions:
$$
v^i={{1}\over{\alpha}}{\left ( {{U^{i}}\over{U^{t}}}+
{\beta}^{i} \right )}.      \eqno(26)
$$

In (18) ${\varepsilon}$ is the total energy density, defined as [5]:
$$
{\varepsilon}{\equiv}{\varepsilon}_{p}+{\varepsilon}_{f},   \eqno(27)
$$
$$
{\varepsilon}_{p}{\equiv}(mn+Pv^2){\Gamma}^2,   \eqno(28)
$$
$$
{\varepsilon}_{f}{\equiv}{{1}\over{8{\pi}}}(E^2+B^2);     \eqno(29)
$$
${\vec S}$ is the total momentum density, defined as:
$$
{\vec S}{\equiv}{\vec S}_{p}+{\vec S}_{f},    \eqno(30)
$$
$$
{\vec S}_{p}{\equiv}(mn+P){\Gamma}^2{\vec v},   \eqno(31)
$$
$$
{\vec s_{f}}{\equiv}{{1}\over{4{\pi}}}({\vec E}{\times}{\vec B}),   \eqno(32)
$$
and $T_{ik}$ is defined as
$$
T_{ik}{\equiv}T_{ik}^{p}+T_{ik}^{f},   \eqno(33)
$$
$$
T_{ik}^{p}{\equiv}(mn+P){\Gamma}^2v_{i}v_{k}+Pg_{ik},   \eqno(34)
$$
$$
T_{ik}^{f}{\equiv}{{1}\over{4{\pi}}}{\left [-(E_{i}E_{k}+B_{i}B_{k})+
{{1}\over{2}}(E^2+B^2)g_{ik} \right]}.    \eqno(35)
$$

Note that the roots of the determinants of $||g_{{\alpha}{\beta}}||$ and
$||{\gamma}_{ik}||$ are equal to:
$$
{\sqrt{g}}{\equiv}(det(g_{{\alpha}{\beta}}))^{1/2}={\Sigma}
sin{\theta},   \eqno(36)
$$
$$
{\sqrt{\gamma}}{\equiv}(det({\gamma}_{ik}))^{1/2}={\Sigma}sin{\theta}/
{\alpha}.   \eqno(37)
$$

\section{Main consideration}

Let ${\bar {\theta}(r)}$ denote the jet outer boundary and

$$
{\epsilon}{\equiv}{\left ({{{\theta}}\over{{\bar
{\theta}}}}\right)}^2,  \eqno(38)
$$
be the dimensionless angular variable. From now on, we shall consider
``narrow jet" i.e., ${\bar {\theta}(r)}$ is assumed to be as small that
$sin{\theta}{\approx}{\theta}$ and $cos{\theta}{\approx}1$. Surely, the
same is true for the angular coordinate ${\theta}<{\bar {\theta}(r)}$ inside
the jet. The assumption noticeably simplifies the angular dependence of
various quantities appearing in the theory. For example, from (37) we get:
$$
{\sqrt{\gamma}}{\approx}{{(r^2+a^2)^{3/2}}\over{{\Delta}^{1/2}}}
{\theta}, \eqno(39)
$$
and owing to the same assumption:
$$
{\alpha}{\approx}{\left({{{\Delta}}\over{r^2+a^2}}\right)}^{1/2}, \eqno(40)
$$
$$
{\beta}^{\phi}{\approx}-{{2aMr}\over{(r^2+a^2)^2}}. \eqno(41)
$$

Let us introduce, now, the magnetic flux function ${\psi}(r,{\theta})$ [8],
defined as to have:
$$
B^{r}={{1}\over{{\sqrt{{\gamma}}}}}{\partial}_{\theta}{\psi}
({\epsilon}),  \eqno(42)
$$
$$
B^{{\theta}}=-{{1}\over{{\sqrt{{\gamma}}}}}{\partial}_{r}{\psi}
({\epsilon}),   \eqno(43)
$$

then it is easy to check out that the "no monopoles" condition (17) is
satisfied automatically, and also
$$
B^{\theta}={\left
[-{{{\partial}{\epsilon}/{\partial}r}\over{{\partial}{\epsilon}/
{\partial}{\theta}}} \right]}B^{r}={\sqrt
{\epsilon}}{{{\partial}{\bar {\theta}(r)}}\over{{\partial}r}}B^{r}.
\eqno(44)
$$

Inserting (44) into the (17) and resolving it for $B^r$ we
get:
$$
{\bar B_{r}}{\equiv}{\sqrt {g_{rr}}}B^{r}={{{\Phi}}\over{(r^2+a^2)
{\bar {\theta}}^2(r)}},   \eqno(45)
$$
where ${\bar B_{r}}$ is the ``physical" radial component of the magnetic
field vector, while the ${\Phi}$ is some constant of integration.

Equations  (15) and (16) may be combined to give the so called
``nduction equation":
$$
rot[({\alpha}{\vec v}-{\vec {\beta}}){\times}{\vec B}]=0.    \eqno(46)
$$

Poloidal components of this equation lead to the simple relation
between poloidal components of ${\vec v}$ and ${\vec B}$:
$$
v^{r}B^{\theta}=v^{\theta}B^{r},  \eqno(47)
$$
which, together with (44) leads to
$$
v^{\theta}={\sqrt{{\epsilon}}}{{{{\partial}{\bar \theta}}(r)}
\over{{\partial}r}}v^{r}. \eqno(48)
$$

The toroidal component of (46), after taking into account of the (17) may
be written as
$$
{\partial}_{r}[{\alpha}{\sqrt{{\gamma}}}v^{r}B^{\phi}]+{\partial}_{\theta}
[{\alpha}{\sqrt{{\gamma}}}v^{\theta}B^{\phi}]=
{\sqrt{{\gamma}}}B^{r}{{{\partial}
{\Omega}}\over{{\partial}r}}+{\sqrt{{\gamma}}}B^{\theta}{{{\partial}
{\Omega}}\over{{\partial}{\theta}}}.  \eqno(49)
$$

In this paper we assume that ${\Omega}={\Omega}(r)$. It allows us to
neglect the last term on the right in (49). Remained equation may
be solved separately as inside [at ${\theta}<{\bar \theta}(r)$] as
outside the jet [when ${\theta}>{\bar {\theta}(r)}$]. In such a way
we get:
$$
B^{\phi}(r)={{{\Phi}({\Omega}-{\Omega}_{0})}\over{(r^2+a^2)v^{r}
{\bar {\theta}}^2(r)}},~~~{\theta}{\le}{\bar {\theta}},  \eqno(50a)
$$
$$
B^{\phi}(r)={{{\Phi}({\Omega}-{\Omega}_{0})}\over{(r^2+a^2)v^{r}
{\theta}^2(r)}},~~~{\theta}>{\bar {\theta}}.  \eqno(50b)
$$

Let us consider, now, the momentum conservation equation (18). For
stationary case it may be rewritten as:
$$
-{{1}\over{{\alpha}}}({\vec {\beta}}{\cdot}{\nabla}){\vec S}={\varepsilon}
{\vec g}+{\bf H}{\cdot}{\vec S}-{{1}\over{{\alpha}}}{\nabla}({\alpha}{\bar
T}).  \eqno(51)
$$

For the ${\phi}$-component of this equation taking into account
that $g_{\phi}=0$ and
$$
{1\over{{\alpha}}}{\beta}^{k}S_{{\phi};k}=H_{{\phi}k}S^{k},
\eqno(52)
$$
we get simple equation:
$$
({\alpha}{\sqrt{{\gamma}}}T^{k}_{\phi})_{,k}=0.   \eqno(53)
$$

For the poloidal components of the same equation, taking into account
that
$$
{1\over{{\alpha}}}{\beta}^{k}S_{i;k}=-H_{{\phi}i}S^{\phi}~~~i=r,{\theta},
\eqno(54)
$$
we derive the following equation
$$
{\varepsilon}g_{i}+(H_{{\phi}i}+H_{i{\phi}})S^{\phi}-{{1}\over{{\alpha}
{\sqrt{{\gamma}}}}}[{\alpha}{\sqrt{{\gamma}}}T^{k}_{i}]_{,k}+{1\over2}
T^{mn}g_{mn,i}=0.
\eqno(55)
$$

Note that for the narrow jet:
$$
g_{r}=-{{M(r^{2}-a^{2})}\over{(r^{2}+a^{2}){\Delta}}},  \eqno(56a)
$$
$$
g_{{\theta}}={{2Mra^{2}{\theta}}\over{(r^{2}+a^{2})^{2}}}.
\eqno(56b)
$$

It is also easy to prove that
$$
(H_{{\phi}i}+H_{i{\phi}})S^{\phi}={{1}\over{{\alpha}}}
S_{\phi}{\beta}^{\phi},_{i}.   \eqno(57)
$$

Energy conservation equation in stationary the case may be written simply
as
$$
{\nabla}({\alpha}^{2}{\vec S})=-{\alpha}^{2}{\sigma}_{ik}T^{ik}.
\eqno(58)
$$

If we consider that
$$
{\sigma}_{ik}T^{ik}=-{{1}\over{{\alpha}}}T_{{\phi}}^{i}{\beta}^{{\phi}},_{i}
{\approx}-{{1}\over{{\alpha}}}T_{{\phi}}^{r}{\beta}^{\phi},_{r},
\eqno(59)
$$
then we can rewrite (58) in the following form:
$$
{\nabla}({\alpha}^2{\vec S})={\alpha}T_{\phi}^{r}{\beta}^{\phi}_{,r}.
\eqno(60)
$$

It must be noted that our equations
contain as ``special-relativistic" effects (Lorentz factors and all
that) as, also, purely gravitational effects related to the curving of
absolute space $({\alpha}{\not=}1)$ and the ``frame dragging"
$({\vec {\beta}}{\not=}0)$. In the present paper, in purpose to
simplify the consideration, we shall assume that in the innermost
region of the jet, where gravitational effects are perceptible,
matter moves with non-relativistic velocity (i.e., $v{\ll}1$ and
${\Gamma}{\approx}1$). Thus we shall deal with the
general-relativistic but slow (``non-relativistic" in the sense of
special relativity) jet. Hereafter, we shall need to integrate some of our
equations over $r=const$ surfaces. According to the general theory
element of such surface is:
$$
d^2{\vec x}={\sqrt {g_{{\phi}{\phi}}g_{{\theta}{\theta}}}}d{\phi}d
{\theta}={\alpha}{\sqrt
{\gamma}}d{\phi}d{\theta}{\approx}(r^{2}+a^{2}){\theta}d{\phi}d{\theta}.
\eqno(61)
$$

First of all, let us integrate continuity equation:
$$
{\partial}_{r}{\left [{\alpha}{\sqrt {\gamma}}{\Gamma}mnv^{r} \right]}+
{\partial}_{{\theta}}{\left [{\alpha}{\sqrt {\gamma}}{\Gamma}mn
v^{{\theta}} \right]}=0.   \eqno(62)
$$

Remembering that ${\Gamma}{\approx}1$ and integrating (62) by
$d^{2}{\vec x}$ we get the following equation:
$$
{\hat M}{\equiv}2{\pi}{\int_{0}^{{\bar {\theta}}}{{\alpha}{\sqrt
{\gamma}}mnv^{r}d{\theta}}}={\pi}(r^2+a^2){\bar {\theta}}^{2}m{\bar
n}v^{r}=const.  \eqno(63)
$$

Note that deriving (63) we made the following assumption about
the angular dependence of $n(r,{\theta})$:
$$
n(r,{\theta})={\bar n}(r)f({\epsilon}),    \eqno(64)
$$
where $f({\epsilon})>0$ is the dimensionless number-density
``profile- function" normalized in such a way as to get:
$$
\int_{0}^{1}f({\epsilon})d{\epsilon}=1.   \eqno(65)
$$

Similarly, under same assumptions, we can integrate toroidal component
of momentum conservation equation (53). We assume that the jet matter
may be treated as the medium with ultrarelativistic temperature
$$
P=nKT,      \eqno(66)
$$
$$
e=mn+({\bar {\gamma}}-1)^{-1}P.    \eqno(67)
$$
where ${\bar {\gamma}}=5/3$. In this case we have:
$$
P+{\varepsilon}=mn{\left (1+{{5KT}\over{2{\mu}}} \right)}.   \eqno(68)
$$

Taking into account (68), after integration of (53) we get:
$$
{{{\hat M}{\xi}}\over{2{\pi}{\alpha}}}{\left (1+{{5KT}\over{2{\mu}}} \right)}
(r^2+a^2)({\beta}^{\phi}+{\Omega})
{\bar {\theta}^2(r)}-{{{\Phi}^2{\alpha}({\Omega}-{\Omega}_{0})}\over{16{\pi}
v^{r}}}={\it L},  \eqno(69)
$$
where ${\it L}$ is some constant of integration and $\xi$ is defined as:
$$
{\xi}{\equiv}{\int_{0}^{1}{\epsilon}f({\epsilon})d{\epsilon}}.
\eqno(70)
$$

Integration of the energy conservation equation also leads to the another
algebraic equation of the following form:
$$
{{\alpha}{\hat M}\over{2{\pi}}}{\left (1+{{5KT}\over{2{\mu}}}
\right)}-{{{\alpha}{\Phi}^{2}({\Omega}-{\Omega}_{0})({\Omega}_{0}-{\omega}})
\over{16{\pi}v^{r}}}={\it L}({\bar {\omega}}-{\omega}),   \eqno(71)
$$
where ${\omega}{\equiv}-{\beta}^{\phi}$ and ${\bar {\omega}}$ is the
another constant of integration.

If we assume that $5KT/2{\mu}{\ll}1$, then (69) can be rewritten as:
$$
{\Omega}={{{\left [{{{\hat M}{\xi}}\over{2{\pi}{\alpha}}}(r^2+a^2){\bar
{\theta}}^2
{\omega}+{\it L}\right]}v^{r}-{{{\alpha}{\Phi}^2}\over{16{\pi}}}{\Omega}_{0}}
\over{{{{\hat M}{\xi}}\over{2{\pi}{\alpha}}}(r^2+a^2){\bar {\theta}^2(r)}v^{r}
-{{{\alpha}{\Phi}^2}\over{16{\pi}}}}}.  \eqno(72)
$$

For the physical jet solution, the numerator and denominator of
(72) must vanish simultaneously at a distance $r=r_{A}$, termed
the Alfven point of the flow. At this point:
$$
v^{r}_{A}={{{\alpha}_{A}^2{\Phi}^2}\over{8{\hat M}{\xi}(r^2_{A}+a^2)
{\bar {\theta}}_A^2}},   \eqno(73)
$$
$$
{\it L}={{{\hat M}m}\over{2{\pi}{\alpha}_{A}}}(r^2_{A}+a^2){\bar {\theta}}_A^2
({\Omega}_{0}-{\omega}_{A})={{{\alpha}_{A}{\Phi}^2({\Omega}_{0}-
{\omega}_{A})}\over{16{\pi}v^{r}_{A}}}.   \eqno(74)
$$

If we neglect similarly term containing temperature in (71), then
it may be soled together with (72) for the angular velocity. In
such a way we get:
$$
{\Omega}={\omega}+{{{\alpha}[{\alpha}+(2{\pi}{\it L}/{\hat
M})({\Omega}_{0}-{\bar {\omega}})]}\over{{\xi}(r^{2}+a^{2}){\bar
{\theta}}^{2}({\Omega}_{0}-{\omega})}}.     \eqno(75)
$$
Boundary condition ${\Omega}(r_{0})={\Omega}_{0}$ implies that
$$
{{2{\pi}{\it L}}\over{{\hat M}}}({\Omega}_{0}-{\bar {\omega}})=-{\alpha}_{0}+
{{{\xi}(r_0^2+a^2){\bar {\theta}}_{0}^{2}({\Omega}_{0}-
{\omega}_0)^2}\over{{\alpha}_{0}}},
$$

and (75) may be rewritten in the following form:
$$
{\Omega}={\omega}+{{{\alpha}}\over{(r^2+a^2)({\Omega}_{0}-{\omega})}}
{\left [{{{\alpha}-{\alpha}_{0}}\over{{\xi}{\bar {\theta}^2}}}+{{(r_{0}^2+a^2)
({\Omega}_{0}-{\omega}_{0})^2}\over{{\alpha}_{0}}}{\left ({{{\bar{\theta}}_0}
\over{{\bar{{\theta}}}}}\right)}^2\right]}.  \eqno(76)
$$

Knowing explicit analytical expression for ${\Omega}(r)$ we can
calculate all other physical quantities connected with it.
In particular, we can get for toroidal magnetic field:
$$
B^{\phi}={\left ({{8{\hat M}{\xi}}\over{{\Phi}}} \right)}{{1}
\over{{\alpha}(r^{2}+a^{2}){\bar {\theta}}^{2}}}{\Biggl[}{{(r^{2}+a^{2})
{\bar {\theta}}^{2}({\Omega}-{\omega})}\over{{\alpha}}}-
$$
$$
{\left (1- {{{\alpha}_{0}^{2}{\epsilon}}\over{\xi}} \right)}{{(r_{0}^{2}
+a^{2}){\bar
{\theta}}_{0}^{2}({\Omega}_{0}-{\omega}_{0})}\over{{\alpha}_{0}}} {\Biggr]}.
\eqno(77)
$$
and then, we can find the expression for radial velocity
$v^{r}$ all other jet physical variables.

It must be emphasized that solutions contain unknown "jet profile"
function ${\bar {\theta}}(r)$. It may be calculated from the radial
component of momentum conservation equation, since it leads to the
first order differential equation for ${\bar {\theta}}(r)$.

\section{Conclusion}

In this letter we demonstrated the general scheme for the construction of the
general-relativistic model of the magnetically driven jet. The method is
based on the usage of the $3+1$ MHD formalism. It is shown that the critical
points of the flow and the explicit radial behavior of the physical
variables may be derived. All jet characteristics may be expressed through
{\it one} quantity: the jet ``profile function" ${\bar {\theta}}(r)$. The
latter quantity may be modelled in some way (i.e., by adopting the
simplest {\it constant open angle jet} approximation, or by using of some
phenomenological jet profile form). Alternative, and more self-consistent,
approach should imply the solution of the complex first order ordinary
differential equation for the ${\bar {\theta}}(r)$ function.

However, the full examination of the problem is beyond the scope of this
work. The results of this letter are of the preliminary character. Full
consideration of the problem is in preparation and will be published
elsewhere.

\section{Acknowledgements}

Our research was supported, in part, by International Science
Foundation (ISF) long-term research grant RVO 300.

\begin{flushleft}
{\Large \bf REFERENCES}
\end{flushleft}
\vskip 0.3cm
\begin{enumerate}
\item
R. D. Blandford, in {\it Astrophysical Jets}, Eds. D. Burgarella, M. Livio
and C. O'Dea, (Cambridge University Press, Cambridge, 1993).
\item
M. C. Begelman, R. D. Blandford \& M. J. Rees, rev. Mod. Phys. {\bf 56},
255, (1984).
\item
K. S. Thorne, \& D. A. MacDonald, Mon. Not. R. Astr. Soc. {\bf 198}, 339,
(1982).
\item
D. A. MacDonald \& K. S. Thorne, Mon. Not. R. Astr. Soc. {\bf 198}, 345,
(1982).
\item
K. S. Thorne, R. H. Price, \& D. A. MacDonald, eds. {\it Black Holes: The
Membrane Paradigm} (Yale University Press, New Haven 1986).
\item
R. V. E. Lovelace, C. Mehanian, C. Mobarry  \& M. E. Sulkanen, Ap.J.S.
{\bf 62}, 1, (1986).
\item
R. V. E. Lovelace, J. C. L. Wang \& M. E. Sulkanen, Ap.J. {\bf 315}, 504,
(1987).
\item
R. V. E. Lovelace, H. L. Berk \& J. Contopoulos, Ap.J. {\bf 394}, 459,
(1991).
\item
J. Contopoulos \& R. V. E. Lovelace, Ap.J. {\bf 429}, 139, (1994).
\item
J. Contopoulos, Ap.J. {\bf 432}, 508, (1994).
\item
J. Contopoulos, Ap.J. {\bf 446}, 67, (1995).

\end{enumerate}

}
\end{document}